\documentclass[aip, cha,amsmath,sd,reprint,amssymb]{revtex4-1}
\usepackage[utf8]{inputenc}
\usepackage{graphicx}
\usepackage{hyperref}
\usepackage{color}
\definecolor{gray}{gray}{0.7}

\graphicspath{{figs/}}

\begin{document}
\title[Quintero-Quiroz et al]{Differentiating resting brain states using ordinal symbolic analysis}
\author{C. Quintero-Quiroz}
\affiliation{Universitat Polit\`ecnica de Catalunya, Departament de F\'isica, Colom 11, 08222 Terrassa, Barcelona, Spain.}%
\author{Luis Montesano}
\affiliation{Bitbrain, Zaragoza, Spain.}%
\author{A. J. Pons}
\author{M. C. Torrent}
\affiliation{Universitat Polit\`ecnica de Catalunya, Departament de F\'isica, Colom 11, 08222 Terrassa, Barcelona, Spain.}%
\author{J. Garc\'ia-Ojalvo}
\affiliation{Department of Experimental and Health Sciences, Universitat Pompeu Fabra, Parc de Recerca Biom\`edica de Barcelona, Barcelona, Spain.}%
\author{C. Masoller}
\affiliation{Universitat Polit\`ecnica de Catalunya, Departament de F\'isica, Colom 11, 08222 Terrassa, Barcelona, Spain.}%
\date{\today}

\keywords{time series analysis, ordinal analysis, EEG, brain dynamics}

\begin{abstract}
Symbolic methods of analysis are valuable tools for investigating complex {time-dependent} signals. 
{In particular,} {t}he ordinal method defines {sequences of} symbols according to the ordering in which values appear in a time series. {This method} has been shown to yield useful information, even when applied to signals with large noise contamination. Here we use ordinal analysis to investigate the transition between eyes closed (EC) and eyes open (EO) resting states. We analyze two {EEG} datasets (with 71 and 109 healthy subjects)  with different recording conditions (sampling rates and number of electrodes in the scalp).
Using as diagnostic tools the permutation entropy, the entropy computed from symbolic transition probabilities, and an asymmetry coefficient (that measures the asymmetry of the likelihood of the transitions between symbols) we show that ordinal analysis applied to the raw data distinguishes the two brain states. In both datasets we find that the EO state is characterized by higher entropies and lower asymmetry coefficient, as compared to the EC state. Our results thus show that these diagnostic tools have potential for detecting and characterizing changes in time-evolving brain states. 
\end{abstract}
\maketitle

\begin{quotation}
In the ``big data'' era, many efforts are being devoted to extracting useful information from complex signals. 
The human brain is one of the most complex systems that one can try to understand. 
In the last decades, the development and popularization of recording techniques such as electroencephalography (EEG), magnetoencephalography (MEG) and functional magnetic resonance imaging (fMRI), have provided the scientific community with a huge amount of data: different types of brain signals, recorded with different spatio-temporal resolution, under different behavioral or cognitive states, from healthy or from dysfunctioning subjects. 
The underlying brain states are, in spite of many efforts, still poorly understood. 
Here we use a symbolic analysis tool to investigate EEG signals recorded from healthy subjects during a simple behavioral task: the subjects remain in resting state with eyes closed (EC state) during an interval of time, and then open their eyes (EO state). 
We show that symbolic analysis applied to the raw EEG signals detects the transition and identifies subtle differences between the EC and EO brain states.
\end{quotation}

\section{Introduction}
Changes in brain states detected through the analysis of electroencephalography (EEG) signals can be used for translating brain signals into operational commands, and in fact, EEG analysis is one of the techniques used for brain-computer interfaces.
 
Several methods have been used to detect underlying changes in the behavior of dynamical systems from observed data, and one of these, ordinal analysis\cite{PhysRevLett.88.174102,cao2004detecting,Masoller2015}, has been demonstrated to be computationally efficient and to perform well even with very noisy data \cite{zanin2012permutation,carlos_njp_2015}. 
Due to these advantages, ordinal analysis has been used in the field of neuroscience, specifically in the area of epilepsy, for detecting, anticipating and characterizing seizures \cite{li2014using,nicolaou2012detection,olofsen2008permutation,bruzzo2008permutation,li2007predictability,veisi2007fast,ren2015real}. 

Since the early 1930's it is well known that alpha waves dominate the EEGs of healthy individuals when they are resting with their eyes closed, and that this activity diminishes when their eyes are opened \cite{Smith1938,Jasper1936,Adrian1934,Berger1933}.
Therefore, a simple method to detect the Eyes-Closed (EC) to Eyes-Open (EO) transition is by using the Fourier spectrum to estimate the difference of the power of the alpha frequency components \cite{Barry2007,Barry2017}.
However, this approach has the drawback of requiring a certain time-window for computing the power spectra.
Another approach to study the EC-EO transition is to use the synchronization likelihood\cite{jin2014preserved} or the mutual information\cite{tan2013difference} to find changes in the functional brain networks that characterize the two brain states. 
However, constructing functional brain networks is computationally demanding, and comparing them is a challenging task because it is not always possible to discriminate reliably between differences that are due to constrains imposed by method of network construction, or due to genuine changes in brain states\cite{van2010comparing,nat_com_2017}.

The aim of this paper is to investigate if the ordinal approach can accurately discriminate between EC and EO brain states. 
In Sec.~\ref{sec:dataset} we describe the datasets analyzed, in Sec.~\ref{sec:methods} we describe the ordinal-pattern methodology and the quantifiers used to characterize the EC and EO states. 
Sec.~\ref{sec:results} presents the results obtained and Sec.~\ref{sec:conclusions} summarizes our conclusions.

\section{Datasets}
\label{sec:dataset}
We use two EEG datasets with different number of subjects and recording conditions, which are summarized in the Table~\ref{tb:1}. 
Dataset one (DTS1) was collected by the Bitbrain company\cite{bitbrain}. 
The EEG signals were recorded from 71 healthy subjects that remained with eyes closed and eyes open during a period of two minutes each. Dataset two (DTS2), which is freely available\cite{BCI2000,PhysioBank}, consists of EEG recordings of 109 subjects performing the same task, in this case for a period of one minute in each of the two states. 

\begin{table}[tb]
\caption{Description of the datasets used.}
\centering
 \begin{tabular}{|c|c|c|}
  \hline
    			& DTS1 		& DTS2	\\
  \hline
    Sampling rate(Hz)	&  256		& 160	\\
  \hline
    Time task(seg)	&  120		& 60	\\
  \hline
    Total points	&  30720	& 9600	\\
  \hline
   Number of electrodes	&  16		& 64	\\
  \hline
    Number of subjects	&  71		& 109	\\
  \hline
 \end{tabular}
\label{tb:1}
\end{table}

\begin{figure*}[!ht]
      \includegraphics[width = 0.25\linewidth]{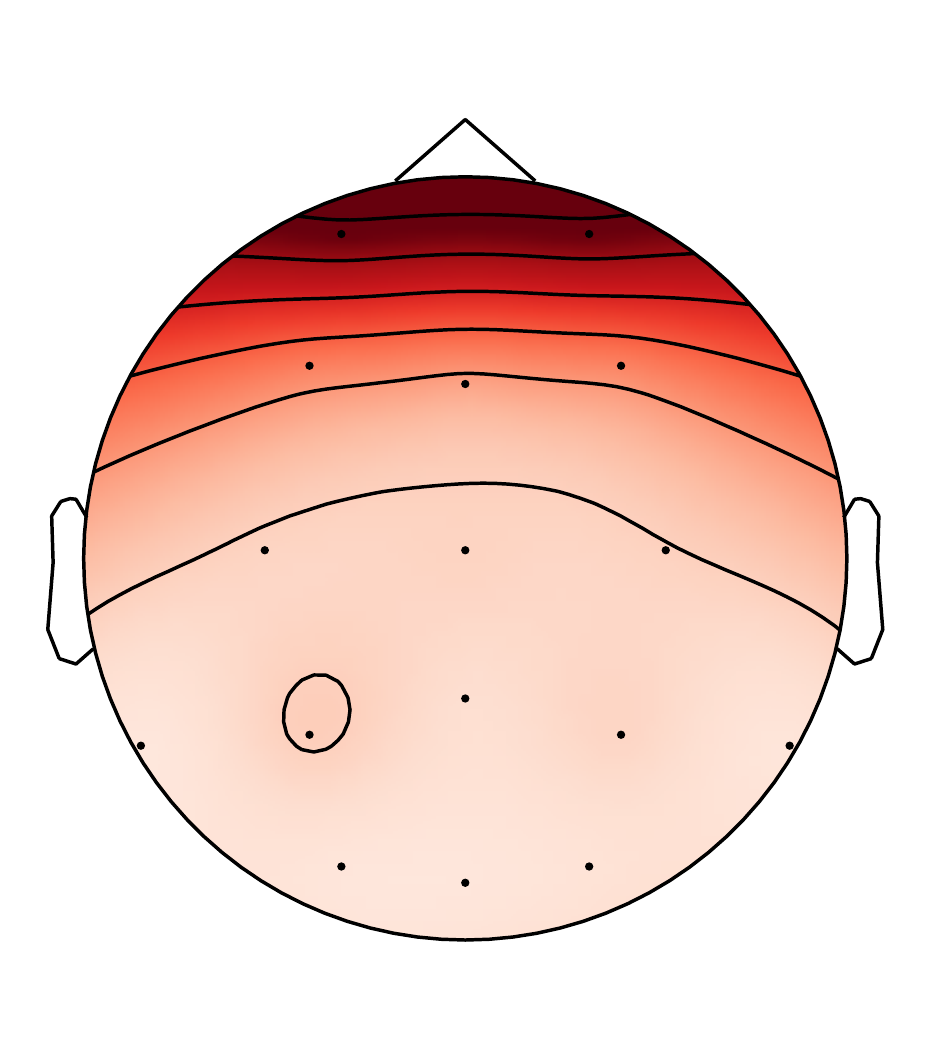}
      \includegraphics[width = 0.7\linewidth]{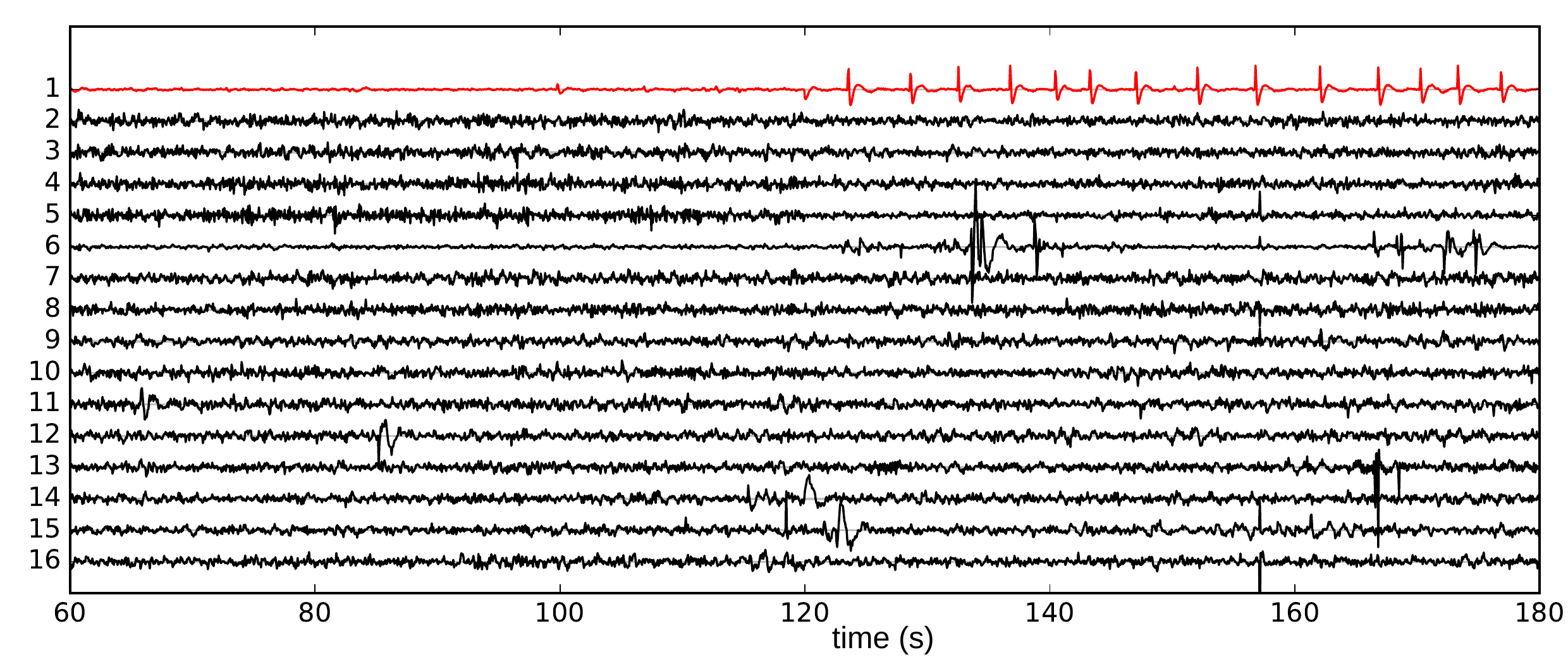}
  \caption{Example of Independent Component Analysis (ICA) related to eye blinks in the EEGs, for a given subject of DTS1. 
  Left: spatial contribution of the selected ICA component, ploted in red in the left panel. 
  Right: all the individual components obtained from the ICA function, in red the component related to the eye blinks, shown in the left panel.}
  \label{fig:icac}
\end{figure*}

We removed the artifacts related to eye blinking following the standard procedure: we applied the Independent Component Analysis (ICA) using the function ICA from the MNE library on Python\cite{gramfort2014mne,gramfort2013meg} and filtered  out the component related to the blinks (see Fig.~\ref{fig:icac}).

\begin{figure*}[!ht]
      \includegraphics[width = 0.25\linewidth]{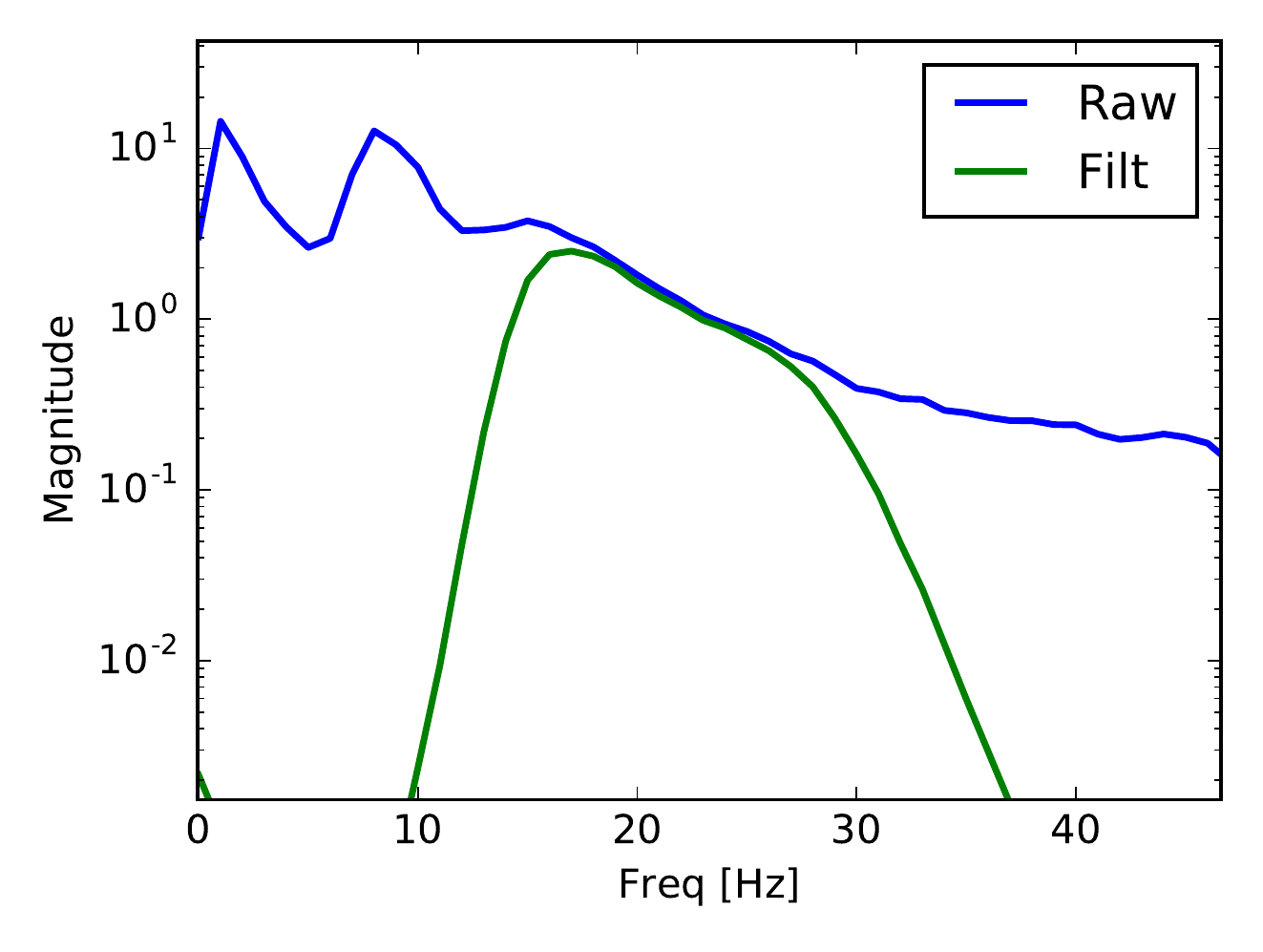}
      \includegraphics[width = 0.7\linewidth]{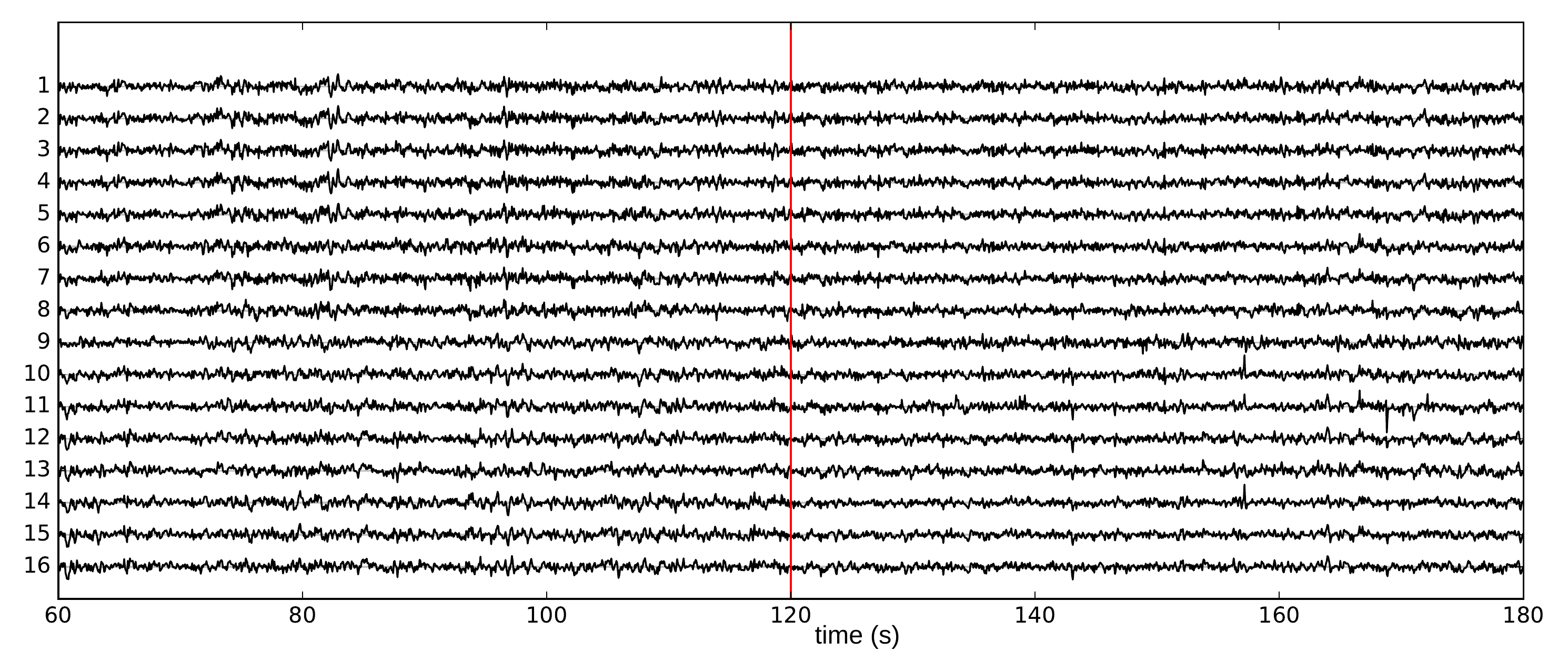}
  \caption{Filtering of the alpha band from the ICA data. 
  The left panel displays the power spectrum of the EEG of a subject before (blue line) and after (green line) filtering. 
  The right top panel displays the post-processed EEG's time-series (after the filtering), the vertical line indicates the time of the eyes closed -- eyes open transition. 
  The subject is the same as in Fig.~\ref{fig:icac}. }
  \label{fig:pow}
\end{figure*}

It is well known that alpha waves are a dominant component in EEG signals during eyes closed conditions, and are reduced when the eyes are open\cite{Berger1933,Barry2007}. 
Therefore, in order to determine whether changes detected through ordinal analysis are only due to the change of the strength of alpha waves, we analyze and compare the results obtained from the \textit{raw} time series, and from \textit{filtered} time series where both eye blinking artifacts and the frequency component of the alpha band were removed (by using a band pass filter between 14 and 31~Hz, see Fig.~\ref{fig:pow}).  

\section{Methods}
\label{sec:methods}
We apply ordinal analysis in non-overlapping windows of 1 second\cite{nicolaou2012detection}, and thus, the number of data points in the window, $w$, is equal to $256$ for DTS1 and to $160$ for DTS2. Then, for each electrode $i$, the time-series, $x_{i}(t)~=~ \{x(1),x(2),\cdots,x(w)\}$ is transformed into a sequence of symbols, $s_{i}(t)$, by using the ordinal rule\cite{PhysRevLett.88.174102,zanin2012permutation}, explained in what follows. 

To define the ordinal patterns we consider vectors of dimension $D$ formed by consecutive data points, i.e. $\{x(j), x(j+1), \cdots, x(j+D-1)\}$, and then assign a symbol according to the ordinal relationship (from the largest to the smallest value) of the $D$ entries in the vectors. For example, with $D = 2$ there are 2 ordinal patterns ($D!$): $x (t_{j} ) < x (t _{j +1})$ corresponding to the ordinal pattern `01’ and $x (t_{i} ) > x (t_{i + 1} )$ corresponding to the ordinal pattern `10’. Then we computed the frequency of occurrence of the D! different patterns in the signal of electrode $i$, and by averaging over all the electrodes, computed the probability of each pattern.

Then, the permutation entropy (PE) is calculated as:
\begin{equation}
 \text{PE} = -\sum_{j} p_{\pi_j} \ln p_{\pi_j}
\label{eq:pe}
\end{equation}
where $p_{\pi_j}$ is the probability of pattern $\pi_j$ along all the electrodes. 
In this way, the PE is a measure of the entropy of the brain EEG signals, in the given time window. 
If the EEG signals are generated by fully random processes, all symbols are equally probable and the PE is maximum, $PE=\ln(D!)$.

Additional diagnostic tools were proposed by \citeauthor{Masoller2015}\cite{Masoller2015}, which are based in the transition probabilities (TPs) between consecutive symbols defined from non-overlapping data values. The transition probability from pattern $\pi_a$ to pattern $\pi_b$ is the relative number of times pattern $\pi_a$ is followed by pattern $\pi_b$, in the sequence $s(t)$:
\begin{equation}
 M_{a, b} = \frac{\sum_{t}^{w-1} N[s(t)=\pi_a,s(t+1)=\pi_b]}{\sum_{t}^{w-1}N[s(t)=\pi_a]}.
\end{equation}
With this definition, the transition probabilities are normalized such that $\sum_{b} M_{ab} = 1$.
Then, exploiting this normalization, an entropy can be associated to the transition probabilities of each pattern as $s_a=-\sum_{b} M_{ab}~\ln M_{ab}$, and its average 
\begin{equation}
s_n   = \frac{\sum_{a} s_a}{D!},
\label{eq:sn}
 \end{equation}
is another measure of the entropy of the EEG signal. 
If a signal is generated by fully random processes, all transition probabilities will be equal and thus, $s_a = \ln(D!)$ for all $\pi_a$, and $s_n = \ln(D!)$. 
In addition, we calculate the transition asymmetry coefficient\cite{Masoller2015}, 
\begin{equation}
a_{c}  =  \frac{\sum_{a} \sum_{b\neq a} \left|M_{ab}-M_{ba}\right|}{\sum_{a} \sum_{b\neq a} \left(M_{ab}+M_{ba}\right)},
\label{eq:ac}
 \end{equation}
which is equal to zero if transition probabilities are fully symmetric ($M_{ab}=M_{ba}$ for all $\pi_a,\pi_b$), and equal to one if they are fully asymmetric (either $M_{ab}=0$ or $M_{ba}=0$, for all $\pi_a,\pi_b$). 
If the EEG signals are generated by fully random processes, then the transition probabilities will be all equal and $a_c=0$.

In the following section the analysis is performed with non overlapping patterns of length $D=4$ (similar results were found with $D=3$). There are $4!=24$ possible patterns and 24$\times$24=576 possible transitions. 
For the dataset DTS1 (DTS2), taking together the 16 (71) electrodes, in each time window the symbolic sequence contains 4048 (10048) patterns and 4032 (9984) transitions.
While the number of patterns is clearly sufficient to compute the probabilities of the 24 patterns with good accuracy, longer sequences are needed to compute the 576 TPs with similar accuracy. Nevertheless, we will show that the TP-based diagnostic tools, $s$ and $a_c$, can also detect changes in datasets. As a measure of statistical significance we calculate the $p$-value using Welch's t-test and consider, as null hypothesis, that the signals represent the same state.

\section{Results}
\label{sec:results}
\begin{figure}[!ht]
 \centering
 \begin{minipage}{\linewidth}
  \includegraphics[width = .8\linewidth]{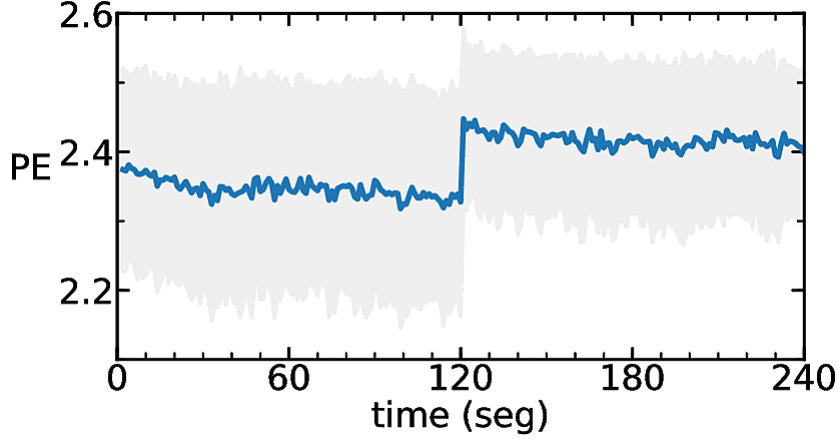}
 \end{minipage}
 \begin{minipage}{\linewidth}
  \includegraphics[width = .8\linewidth]{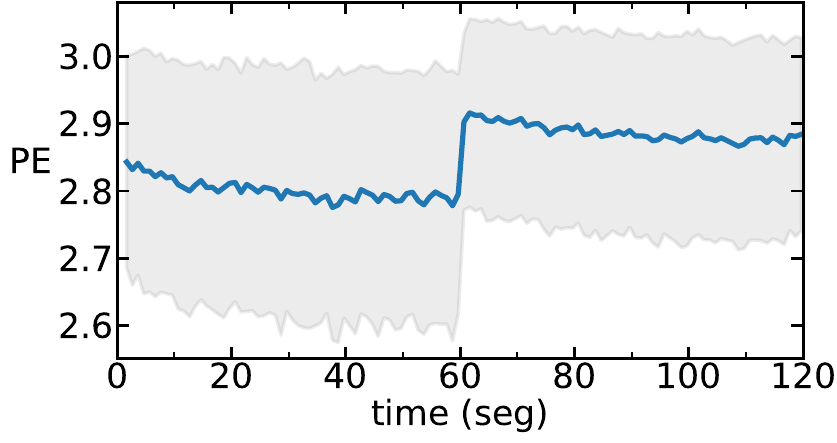}
 \end{minipage}
 \caption{Permutation entropy, Eq.~\ref{eq:pe}, from raw time series of DTS1 (top) and DTS2 (bottom). 
 In DTS1 the subjects open their eyes at 120~s; in DTS2, at 60~s. 
 The blue line indicates the mean value of the PE for all the subjects, and the shaded area indicates one standard deviation of the PE values. }
 \label{fig:pe0}
\end{figure}

We begin by calculating the PE for the raw, unfiltered time series. 
Figure~\ref{fig:pe0} displays the results obtained from DTS1 and DTS2 and we can see that there is a significant difference between the PE values of the eyes-closed and eyes-open states. 
The entropy is computed for each subject, and then is averaged over all the subjects (71 or 109, depending on the dataset). The shaded area represents one standard deviation of the PE values of all subjects, and we note that there is large variability, thus, the PE value does not allow a full discrimination between the two states. We note that the average PE value is slightly different for the two datasets, which is attributed to the fact that they have different spatial and temporal resolution. We also note that the average value of the PE is significantly different from the maximum possible value (which occurs when the patterns are equally probable, and for $D=4$, PE$_{\max}=\ln 24=3.18$). This reveals the presence of patterns with high and low frequency of occurrence in the symbolic sequence. An inspection of the ordinal probabilities reveals that the ``trend'' patterns (generated by consecutively increasing or decreasing data values) are more expressed, in all the channels and for both, EO and EC states. For future work, it would be interesting to investigate if the probabilities of certain patterns (which could be defined by using a lag, as in\cite{parlitz}), allow for a better discrimination of the two states.    


\begin{figure}[!ht]
 \centering
 \begin{minipage}{\linewidth}
  \includegraphics[width = .8\linewidth]{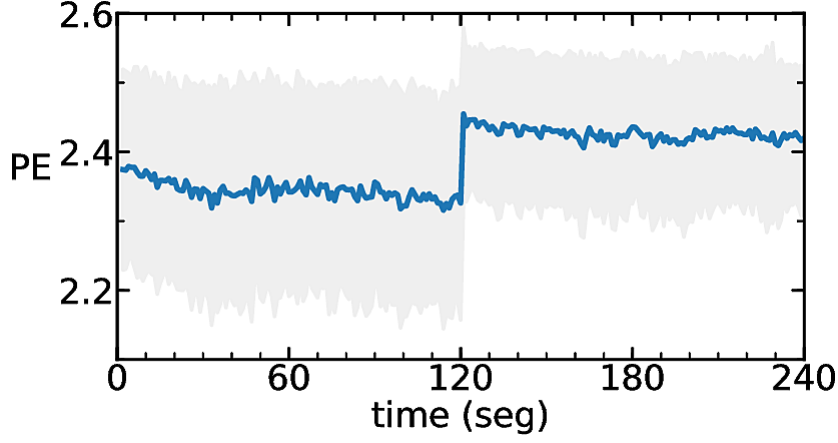}
 \end{minipage}
 \begin{minipage}{\linewidth}
  \includegraphics[width = .8\linewidth]{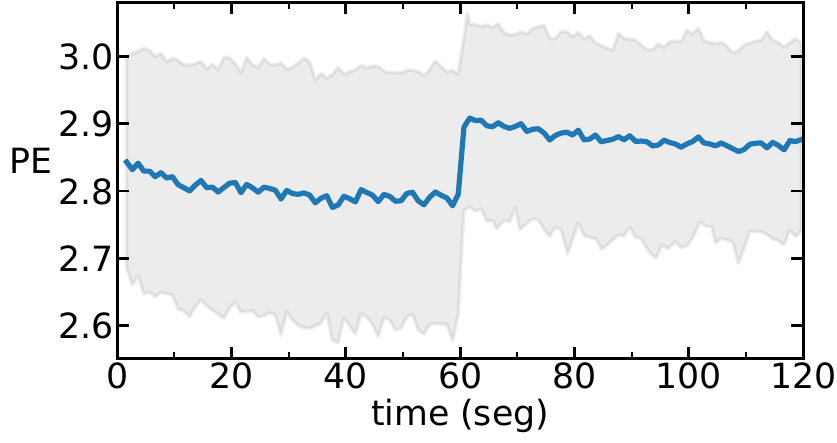}
 \end{minipage}
 \caption{Permutation entropy, Eq.~\ref{eq:pe},
 computed from filtered time series of DTS1 (top) and DTS2 (bottom).}
 \label{fig:pef}
\end{figure}

Comparing with the results obtained from the filtered time series, displayed in Fig.~\ref{fig:pef}, we note that the PE values remain almost unchanged, which suggests that the PE captures changes in brain dynamics which are not due to the change in the strength of alpha oscillations during the EC-EO transition.

\begin{figure}[ht]
 \centering
   \includegraphics[width = 0.9\linewidth]{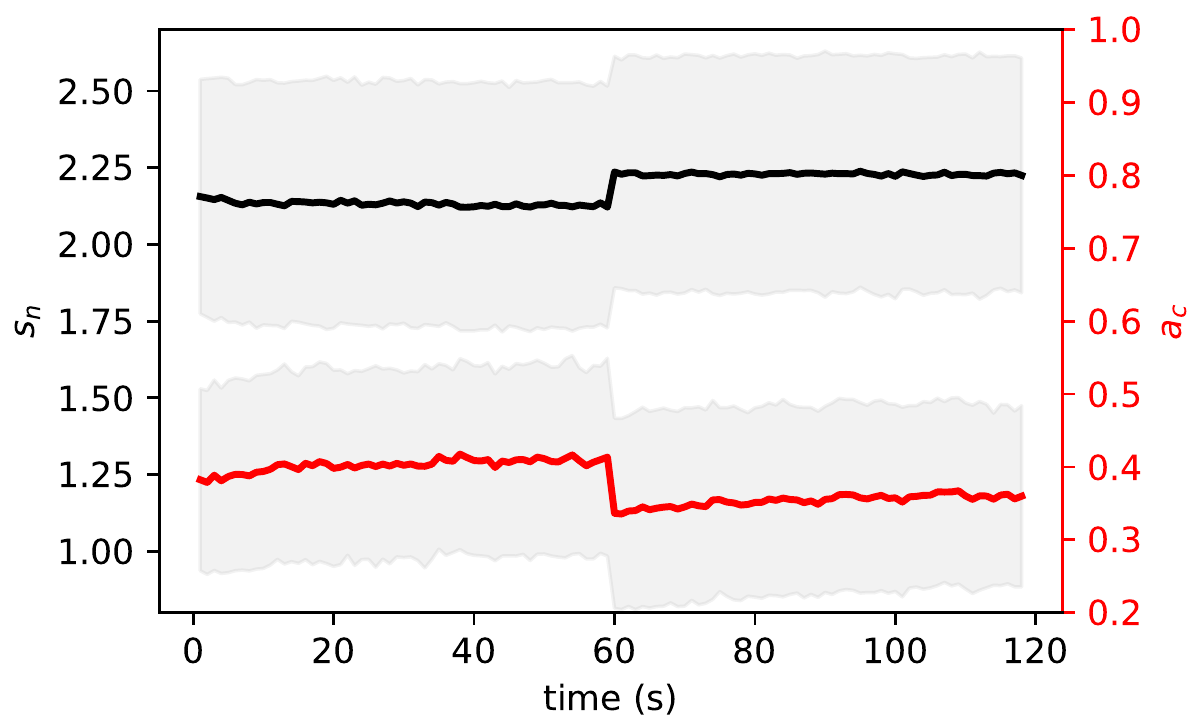}
 \caption{The entropy defined from the transition probabilities, Eq.~\ref{eq:sn}, and the asymmetry coefficient, Eq.~\ref{eq:ac}, computed for the DTS2.}
 \label{fig:net}
\end{figure}

Figure~\ref{fig:net} displays the TP-based measures, $s$ and $a_c$, using the DST2 (similar results were found in DST1), although there is a clear transition around 60s in the mean values, the dispersion in the values of the different subjects is higher than in the PE analysis (likely due to the limited length of the time series, which does not allow a precise estimation of the TPs).

\begin{figure*}[ht]
 \centering
   \includegraphics[width = .8\linewidth]{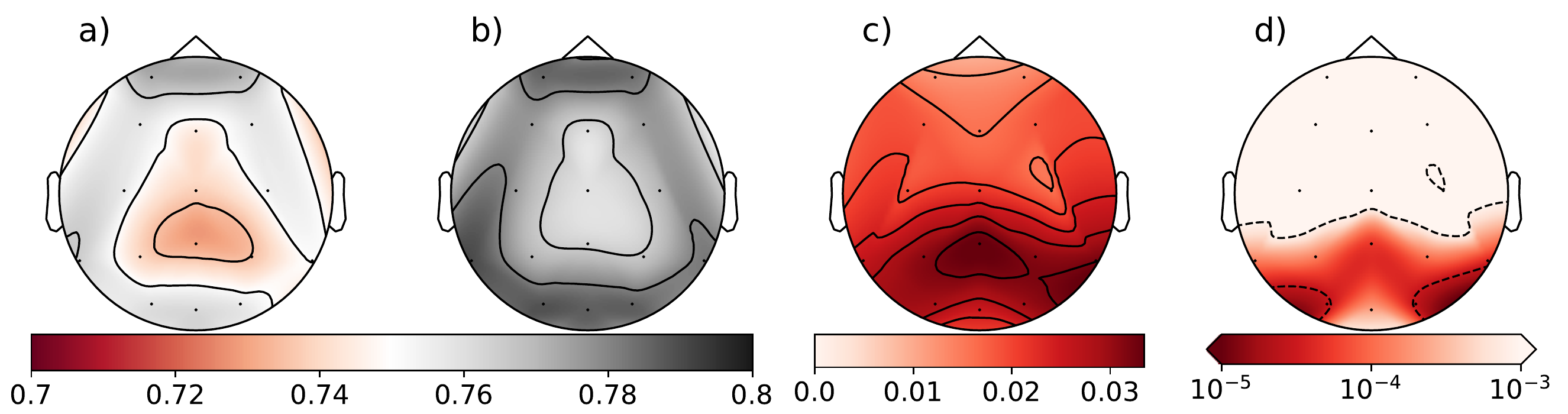}
   \includegraphics[width = .8\linewidth]{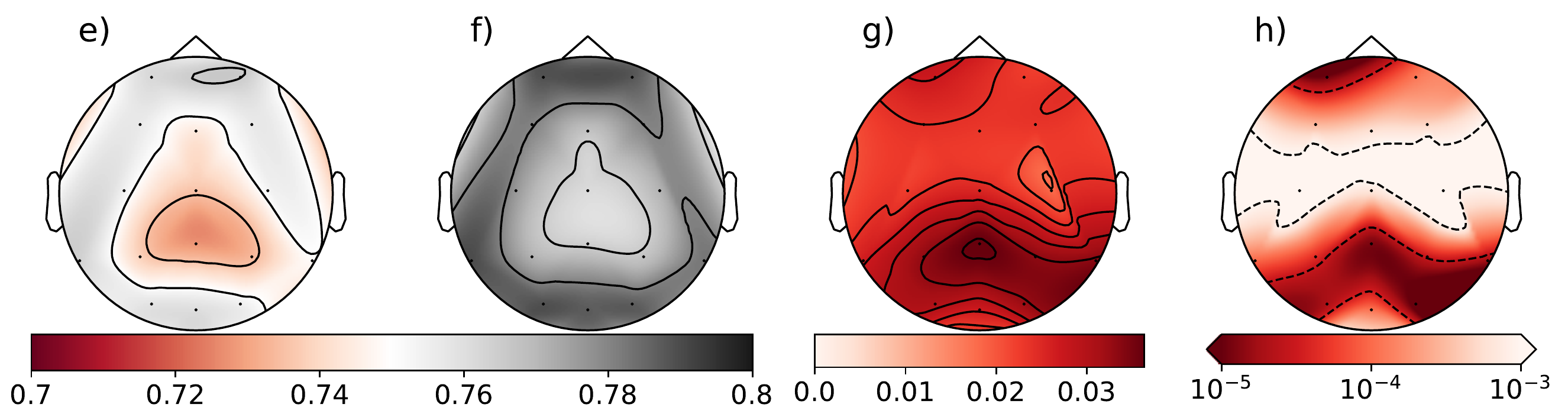}
 \caption{Topographic visualization of the analysis of the raw (top row) and
filtered (lower row) EEG signals of DTS1, average over the subjects. 
Panels a) and e) display the permutation entropy for EC conditions; b) and f) for EO conditions; c) and g) display the difference of the PE values; d) and h) display the $p$ value.}
 \label{fig:brain_mean_dts1}
\end{figure*}

\begin{figure*}[!ht]
 \centering
   \includegraphics[width = .8\linewidth]{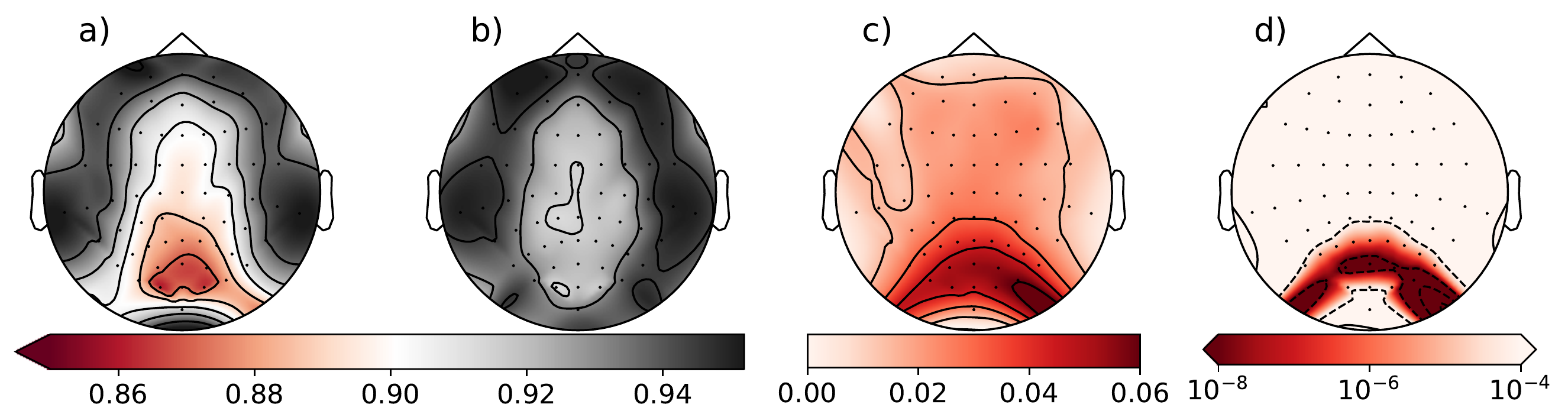}
   \includegraphics[width = .8\linewidth]{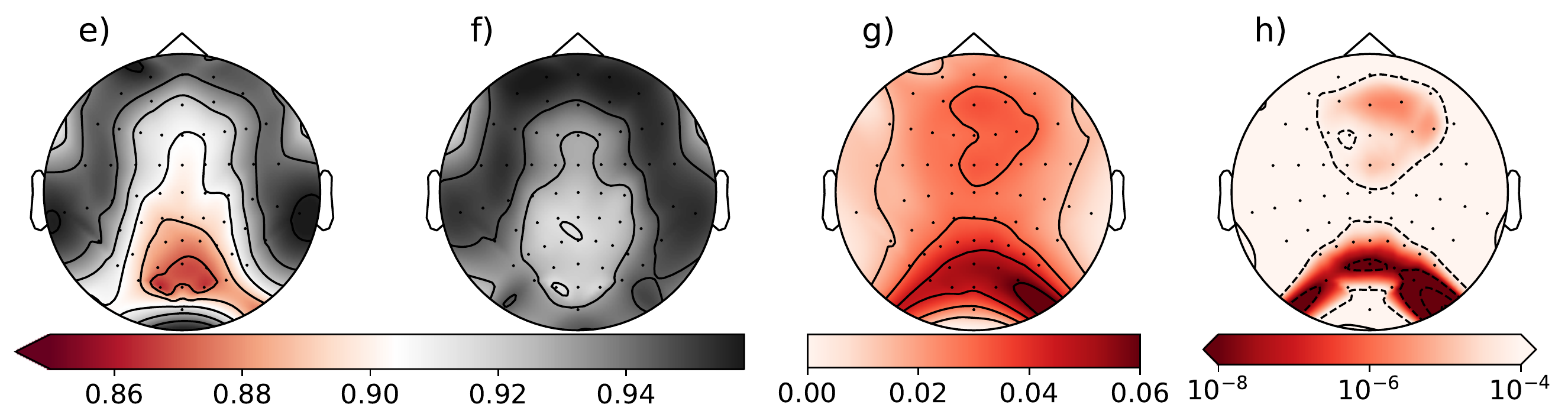}
 \caption{As in Fig.~\ref{fig:brain_mean_dts1}, but for the dataset DTS2.}
 \label{fig:brain_mean_dts2}
\end{figure*}

In Fig.~\ref{fig:brain_mean_dts1} (for DTS1) and Fig.~\ref{fig:brain_mean_dts2} (for DTS2) we present the topographic visualization for the different electrodes, of the PE value averaged over all the subjects, for the EC and EO conditions.
We also present the difference of PE values (PE-open - PE-closed), and the $p$-value.
The results are consistent for the two datasets, the discrepancies are due to their different spatial resolution. 
We also note the low p-values obtained, which confirm the significance of the uncovered differences.

\section{Conclusions}
\label{sec:conclusions}
We have used ordinal time series analysis to investigate EEG signals recorded under eyes-closed (EC) and eyes-open (EO) resting conditions.
We have analyzed two datasets with different spatial and temporal resolutions, and contrasted the results of the analysis of raw time series and filtered time series (where eye blinking artifacts were removed and the alpha frequency band was filtered out).
We used three diagnostic measures, the permutation entropy, PE, which is computed from the probabilities of the ordinal patterns, and two measures, the transition entropy and the asymmetry coefficient, which are computed from the transition probabilities between patterns.

We have found, in both datasets, that the EO state is characterized by higher entropy values, accompanied by a lower asymmetry coefficient, with respect to the EC state. We have also identified which brain regions are more important for distinguishing the two states. No significant difference was detected between the raw data and the pre-processed data, which suggests that the ordinal method can be directly applied to EEG signals, avoiding the need of data pre-processing. Thus, ordinal analysis can be a computationally efficient tool, which could provide extra valuable information for new brain-computer interface protocols.

\section*{Acknowledgments}
This work was supported in part by ITN NETT (FP7 289146), the Spanish MINECO (FIS2015-66503 and FIS2015-66503-C3-2-P) and the program ICREA ACADEMIA of Generalitat de Catalunya.

\bibliographystyle{aipnum4-1}
\bibliography{ref}

\end{document}